\documentclass[sigconf,nonacm]{acmart}

\AtBeginDocument{%
  }

\begin{document}

\title{FuturePrism: Supporting Adolescence in Collaborative Storytelling to Cope with Future Uncertainty}









\author{Yonglin Chen}
\orcid{0009-0004-3081-1813}
\affiliation{%
\institution{School of Design,}
  \institution{Southern University of Science and Technology}
  \city{Shenzhen}
  \country{China}
}
\email{12531641@mail.sustech.edu.cn}

\author{Pengcheng An}
\orcid{}
\affiliation{%
\institution{School of Design,}
  \institution{Southern University of Science and Technology}
  \city{Shenzhen}
  \country{China}
}
\email{anpc@sustech.edu.cn}

\author{Xueliang Li}
\orcid{}
\authornote{Corresponding author.}
\affiliation{%
\institution{School of Design,}
  \institution{Southern University of Science and Technology}
  \city{Shenzhen}
  \country{China}
}
\email{lixl6@sustech.edu.cn}

\renewcommand{\shortauthors}{Trovato et al.}

\begin{abstract}
  FuturePrism is a GenAI-empowered collaborative storytelling system designed to scaffold adolescents to navigate future life challenges. Adolescents often suffer from anxiety related to future uncertainty for lacking the executive function to develop concrete pathways. Operationalizing Snyder’s Hope Theory, the system utilizes a triadic role-play mechanics to externalize cognitive processes through four narrative chapters: The Goal, The Opportunity, The Challenge, and The Agency. An evaluation workshop with 20 adolescents demonstrated that FuturePrism significantly enhances momentary hope levels, particularly in the Agency dimension. Participants reported high levels of narrative immersion and positive feedback towards system usability. Participants also confirmed that the AI-scaffolded collaborative storytelling empowered them to develop positive attitudes towards future challenges.


\end{abstract}




\begin{teaserfigure}
  \includegraphics[width=\textwidth]{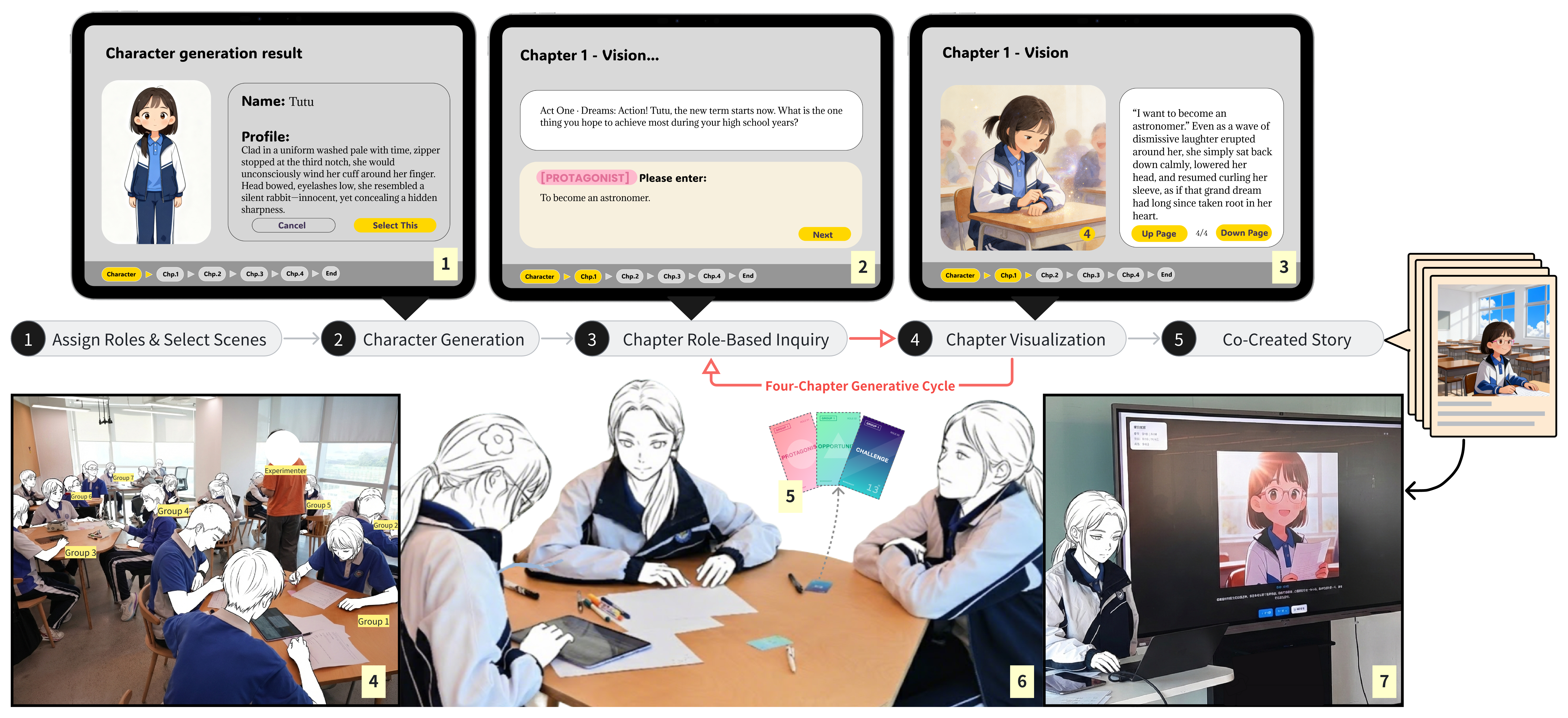}
  \caption{System overview of \textit{FuturePrism} and the collaborative storytelling workshop. The top section illustrates the four-chapter framework of the interaction design-based interface: (1) character generation result, showing the AI-generated character's avatar and profile based on user self-portraits; (2) collecting inputs from assigned roles through interactive questioning; and (3) presenting the AI-generated narrative text and corresponding illustrations based on user inputs. The bottom section depicts the deployment of the system in a workshop: (4) adolescents discussing in groups; (5) the role cards;(6) collaborative interaction using the role cards assigned to "Protagonist," "Opportunity," and "Challenge";  and (7) the presentation session where groups take turns pitching their co-created stories. }
  \Description{}
  \label{fig:teaser}
\end{teaserfigure}

\received{20 February 2007}
\received[revised]{12 March 2009}
\received[accepted]{5 June 2009}

\maketitle

\section{Introduction}

Adolescents aged 13–14 enter Piaget’s formal operational stage, gaining the cognitive capacity to imagine abstract futures \cite{piaget1976piaget}. However, their executive functions—specifically strategic planning—often lag behind, creating a gap between their aspirations and the practical steps required to achieve them \cite{steinberg2017social}. Theoretically, this challenge can be framed in terms of Snyder’s \textit{Hope Theory}, which highlights a disconnect between high-level \textit{Goals} and the concrete \textit{Pathways} needed to achieve them \cite{snyder2002hope}. This inability to construct pathways often fosters uncertainty and anxiety regarding potential life trajectories \cite{nurmi1991adolescents}, further impairing \textit{episodic cognition}—the ability to mentally simulate future events and plan effectively \cite{tang2023links}. There have been various educational methods to scaffold adolescence to overcome anxiety related to future uncertainty, ranging from static methods like questionnaires and lectures to experiential activities, such as educational drama \cite{keller2022factors, tang2024young, lehtonen2012future}. Recent HCI research has increasingly explored immersive technologies, such as Virtual Reality, to facilitate future thinking in safe, simulated environments \cite{zheng2024enhancing, habak2021edge}. However, despite these technological advancements, there remains a notable scarcity of research leveraging Generative AI (GenAI) to dynamically scaffold this complex cognitive process. 



GenAI is uniquely positioned to bridge this gap. Unlike static questionnaires or pre-scripted VR scenarios, GenAI-driven storytelling is increasingly recognized for its ability to dynamically generate text and visuals, thereby creating safe, simulated environments for rehearsing real-world scenarios \cite{hedderich2024piece, fan2025words}. Yet, current applications face two critical limitations in helping adolescents cope with psychological challenges related to future uncertainty. First, most systems focus on creative thinking and skill training (e.g., literacy) \cite{ye2025colin, chen2025characterizing, lehtonen2012future}, neglecting adolescents' individualized psychological needs. Second, these applications often position GenAI as the primary storyteller. This may lead to over-reliance on AI and induce ``cognitive offloading'' \cite{gerlich2025ai}, where users become passive recipients rather than active creators, hindering the development of proactive coping strategies in similar situations.


To address these gaps, we introduce \textit{FuturePrism}, a collaborative storytelling system designed to scaffold adolescent future thinking. We operationalize Hope Theory into a triadic, four-chapter narrative framework. Instead of solitary planning, the system leverages peer dynamics in role-play activities to externalize internal cognitive processes: users adopt three distinct roles—Protagonist, Opportunity, and Challenge—to collaboratively simulate the friction between aspirations and obstacles. Taking the roles during their interaction with the system, users are guided through a collaborative storytelling game that embeds Hope Theory’s components: the Protagonist embodies the Agency to pursue \textit{Goals}, while the interplay between Opportunity and Challenge externalizes the dynamic construction of \textit{Pathways}. Here, GenAI functions not as a mere content generator, but as a narrative scaffold, soliciting high-level decisions to support \textit{Agency} without replacing human imagination.


We evaluated the system through an evaluation workshop involving 20 adolescents (11 males, 9 females; $M_\text{age} = 13.1$). This research makes the following contributions:

\begin{itemize}
    \item \textbf{System Design:} We introduce \textit{FuturePrism}, a GenAI-empowered collaborative storytelling system that operationalizes Hope Theory into a triadic role-play framework. It transforms abstract future planning into concrete social interactions to scaffold adolescents' future thinking and mitigate uncertainty.
    \item \textbf{Empirical Evidence:} We provide empirical findings from a user study ($N=20$) demonstrating that the system significantly enhances momentary adolescents' hope levels (measured through Children's Hope Scale), particularly in the dimension of Agency, while fostering high narrative immersion.
\end{itemize}

\section{System Design of \textit{FuturePrism}}

We developed \textit{FuturePrism}, a tablet-based application designed to scaffold adolescents' future thinking through collaborative role-play. Below, we introduce the system design in terms of its underlying \textbf{Narrative Framework} (structural rules grounded in Hope Theory) and \textbf{Interaction Workflow} (the sequential user journey).





\subsection{A Four-Chapter Narrative Framework} 
\label{sec:narrative_framework}

\textit{FuturePrism} features a narrative framework that guides the AI's generative output. This framework aims to operationalize Snyder's Hope Theory \cite{snyder2002hope} by leveraging GenAI's personalized, flexible, real-time generation capabilities. Diverging from the open-ended creation typical of many GenAI applications, the storytelling process is scaffolded into four distinct chapters. This design not only translates the cognitive elements of hope into narrative plot progressions but also aligns with the classic dramatic arc (Exposition, Rising Action, Climax, and Resolution) \cite{freytag1908freytag}.

\begin{figure*}[htbp] 
    \centering 
    \includegraphics[width=1.0\linewidth]{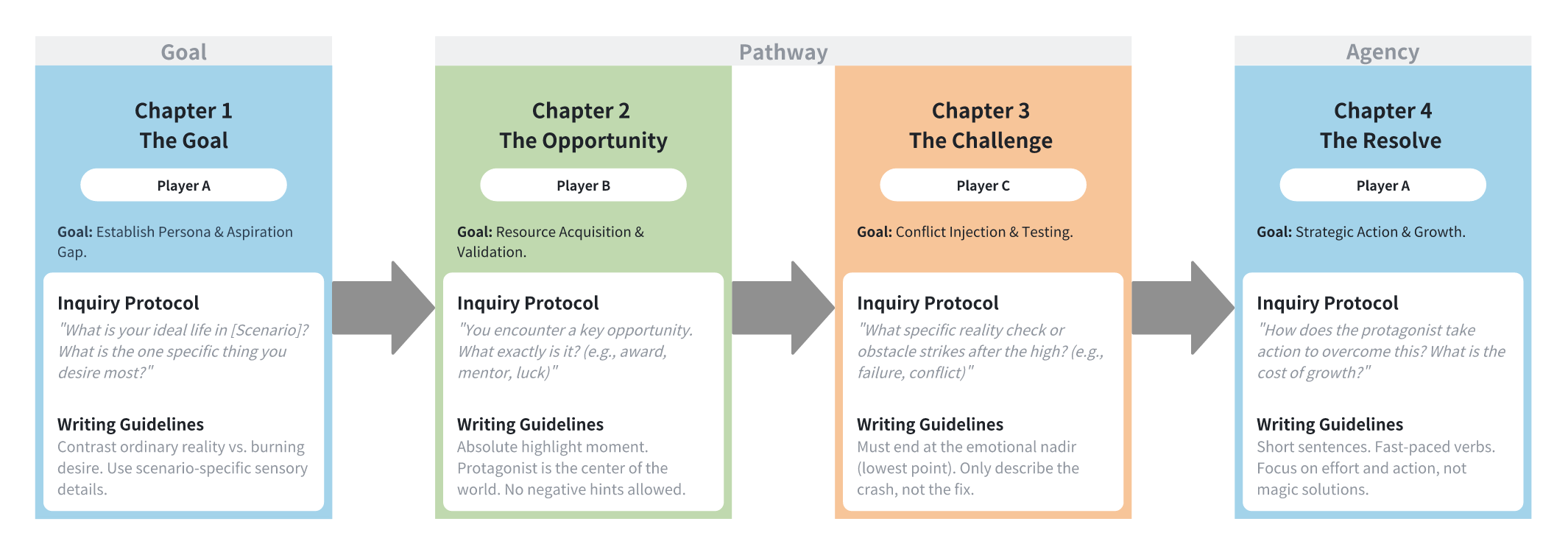} 
    
    \caption{The four-chapter narrative framework of \textit{FuturePrism} that operationalizes Snyder’s Hope Theory (top bar) into a four-chapter role-play structure. Chapter 1 anchors the \textit{Goal}; Chapters 2 and 3 constitute the \textit{Pathways} process; Chapter 4 activates \textit{Agency}, requiring the protagonist to formulate a strategic resolution. The bottom section details the inquiry protocols and writing guidelines used to scaffold the GenAI's output.}
    
    \label{fig:story_framework} 
\end{figure*}

As illustrated in Figure \ref{fig:story_framework}, the collaborative narrative unfolds through a structured four-stage story designed to operationalize the core components of Snyder's Hope Theory. The process begins with \textbf{Chapter 1 (The Goal)}, where Player A articulates a specific aspiration, explicitly anchoring the \textit{Goals} component. The narrative then shifts to \textbf{Chapter 2 (The Opportunity)}, as Player B introduces a positive resource, representing favorable uncertainty to initiate the construction of \textit{Pathways}. Subsequently, \textbf{Chapter 3 (The Challenge)} sees Player C introduce a complication, representing adverse uncertainty that tests the viability of these \textit{Pathways}. Finally, the sequence concludes with \textbf{Chapter 4 (The Resolve)}, where Player A formulates a strategy to overcome the established obstacle, thereby activating the personal \textit{Agency} required to pursue the intended future.


\subsection{Interaction Workflow}
\label{sec:interaction_workflow}


\begin{figure*}[htbp] 
    \centering 
    \includegraphics[width=1.0\linewidth]{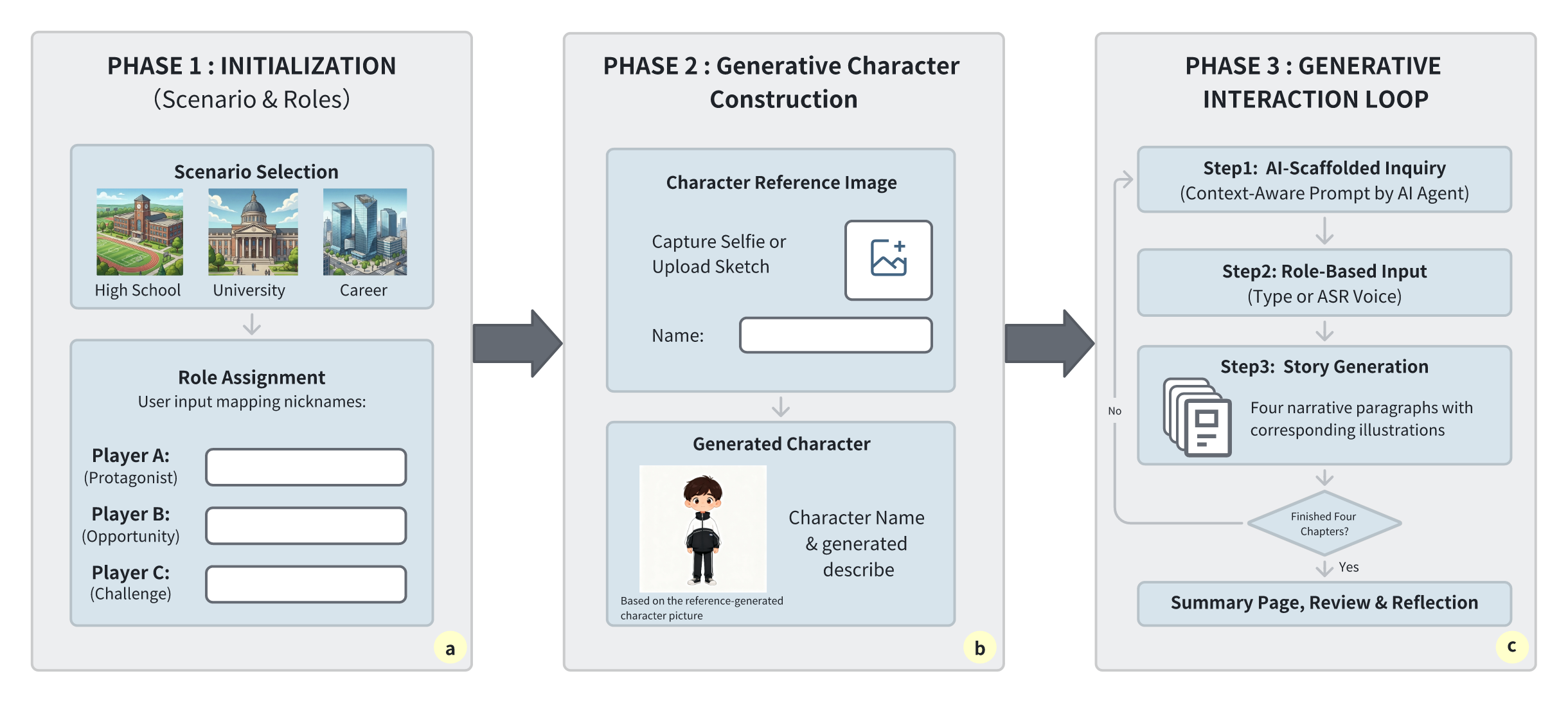} 
    
    \caption{The \textit{FuturePrism} user interface flow: (a) Collaborative role assignment mapping users to narrative perspectives; (b) Character construction using multimodal input (camera/sketch) and Img2Img generation; (c) The main storytelling interface featuring the standardized scaffolding-generation loop.}
    
    \label{fig:system_design} 
\end{figure*}

\subsubsection{Collaborative Setup and Role Allocation}

The session initiates with a collaborative topic selection. The group negotiates to select a specific future life stage to simulate. For this study, we curated three distinct scenarios: \textit{High School Life}, \textit{University Life}, and \textit{Early Career}. These were selected as they represent the most salient and foreseeable developmental milestones for adolescents. Upon confirmation, the interface transitions to the ``Role Assignment'' screen. At this stage, users physically map their real identities to the three narrative perspectives (Protagonist, Opportunity, and Challenge). This step ensures each member understands their specific turn-taking responsibilities as defined in the framework structure (see Figure \ref{fig:system_design}-a).


\subsubsection{Generative Character Construction}
Following scenario selection, the system proceeds to the Character Construction interface (Figure \ref{fig:system_design}-b). To foster player identification and immersion through visual ownership \cite{zarei2020investigating}, the interface allows participants to generate a custom avatar from a real-world selfie or hand-drawn sketch. An image-to-image model then style-transfers this input into a unified illustrative aesthetic, preserving key facial or structural features. Finally, the group reaches a consensus on the protagonist's name, establishing the definitive visual and nominal identity that will anchor the unfolding narrative. A representative example of the character result interface is shown in Figure \ref{fig:teaser}-1.


\subsubsection{The Generative Interaction Loop}

With the character established, the narrative progresses sequentially through the four chapters. Each chapter adheres to a standardized ``Scaffolding-Generation'' loop, strictly guided by the specific inquiry protocols and goals detailed in Figure \ref{fig:story_framework}.

As illustrated in Figure \ref{fig:system_design}-c, the cycle initiates with an \textbf{AI-Scaffolded Inquiry}, where the AI narrator presents a context-aware prompt aligned with the current chapter's objective. For instance, in Chapter 1 (\textit{High School} scenario), the system asks: \textit{``[Protagonist Name] has just stepped into high school... what is the one specific thing they desire most?''} The designated player then provides \textbf{Role-Based Input}; to lower expression barriers, the system supports multimodal entry via text or voice. Upon submission, the \textbf{Story Generation} phase processes this input (approx. 40s latency), expanding the raw ideas into a coherent four-paragraph narrative accompanied by four illustrations that maintain the protagonist's visual consistency. The interface of these generation steps is shown in Figure \ref{fig:teaser}-2 and Figure \ref{fig:teaser}-3.




\subsection{Technical Implementation} 

The generative pipeline is orchestrated via n8n\footnote{\url{https://n8n.io}}, integrating DeepSeek V3.1\footnote{\url{https://api-docs.deepseek.com/news/news250821}} for narrative generation and Seedream 4.0\footnote{\url{https://seed.bytedance.com/en/seedream4_0}} for visual synthesis. To operationalize the scaffolding described in Section \ref{sec:narrative_framework}, we employed a multi-agent architecture where every chapter is mapped to a pre-defined JSON-based prompt schema (encompassing narrative arcs, inquiry protocols, and stylistic guidelines). This structure coordinates three specialized agents—the \textbf{Questioning Agent}, the \textbf{Writing Agent}, and the \textbf{Drawing Agent}—to drive the interaction loop. 

\section{Evaluation Workshop}

To evaluate the efficacy of \textit{FuturePrism}, we conducted an in-person workshop in a classroom at the authors' university. This workshop aimed to assess the system's impact on adolescents' hope levels, narrative engagement, and system usability, and to gather qualitative feedback from participants.


\subsection{Participants}
We recruited 20 adolescents (11 males, 9 females; $M_{age}=13.1, SD=0.3$) from Grade 8 at a local middle school. All participants possessed basic digital literacy. Although their prior experience with Generative AI varied (see Table \ref{tab:participants_split}), notably, only one participant had never utilized such AI tools, thereby minimizing potential novelty bias. Participants were randomly assigned to seven triadic groups (G1--G7). To accommodate the sample size ($N=20$), Group 7 included one volunteer researcher playing a non-intrusive supporting role to complete the triadic interaction mechanism.


\begin{table*}[t]
\centering
\caption{Participant Demographics and Group Composition ($N=20$). IDs follow the format \textit{Group-Role}. }

\label{tab:participants_split}
\begin{tabular}{@{}lcccll || lcccll@{}}
\toprule
\textbf{Grp} & \textbf{ID} & \textbf{Role} & \textbf{Age} & \textbf{Gen.} & \textbf{AI Freq.} & 
\textbf{Grp} & \textbf{ID} & \textbf{Role} & \textbf{Age} & \textbf{Gen.} & \textbf{AI Freq.} \\ 
\midrule
G1 & 1-1 & Protagonist & 13 & M & Monthly & G5 & 5-1 & Protagonist & 13 & M & Weekly \\
   & 1-2 & Opportunity & 13 & M & Weekly   &    & 5-2 & Opportunity & 13 & F & Weekly \\
   & 1-3 & Challenge   & 13 & M & Weekly   &    & 5-3 & Challenge   & 13 & F & Monthly \\
\cmidrule{1-6} \cmidrule{7-12}

G2 & 2-1 & Protagonist & 13 & M & Weekly   & G6 & 6-1 & Protagonist & 13 & M & Monthly \\
   & 2-2 & Opportunity & 13 & M & Weekly   &    & 6-2 & Opportunity & 13 & M & Never \\
   & 2-3 & Challenge   & 13 & M & Monthly &    & 6-3 & Challenge   & 14 & M & Weekly \\
\cmidrule{1-6} \cmidrule{7-12}

G3 & 3-1 & Protagonist & 13 & F & Weekly   & G7 & 7-1 & Protagonist & 13 & F & Weekly \\
   & 3-2 & Opportunity & 13 & F & Daily      &    & 7-2 & Opportunity & 14 & M & Weekly \\
   & 3-3 & Challenge   & 13 & F & Weekly   &    & --  & Challenge   & -- & --& (Volunteer)\\
\cmidrule{1-6} \cmidrule{7-12}

G4 & 4-1 & Protagonist & 13 & F & Daily      &    &     &             &    &   &  \\
   & 4-2 & Opportunity & 13 & F & Daily      &    &     &             &    &   &  \\
   & 4-3 & Challenge   & 13 & F & Weekly   &    &     &             &    &   &  \\
\bottomrule
\end{tabular}
\end{table*}

\subsection{Procedure}

The 90-minute workshop was facilitated by two researchers in a university setting (Figure \ref{fig:teaser}-4). The session unfolded in three phases: \textbf{Phase 1: Orientation (15 mins).} After a brief introduction, participants formed groups and were randomly assigned roles (Protagonist, Opportunity, Challenge) using physical role cards (Figure \ref{fig:teaser}-5). They then completed the \textbf{pre-test questionnaire} (demographics and CHS). \textbf{Phase 2: Collaborative Storytelling (60 mins).} Groups engaged with the \textit{FuturePrism} system following the interaction workflow described in Section \ref{sec:interaction_workflow}. Upon selecting a life scenario and generating their character avatars, groups progressed through the four-chapter narrative cycle (Section \ref{sec:narrative_framework}). Researchers remained on standby to assist with system usage but did not intervene in the creative process. \textbf{Phase 3: Reflection (15 mins).} Groups presented 2-minute story pitches (Figure \ref{fig:teaser}-7). Subsequently, participants completed the \textbf{post-test} (including the Children’s Hope Scale, Narrative Transportation scale, UMUX-Lite, and open-ended questions on subjective experience), concluding the workshop.


\subsection{Measures}
Aligned with the procedure described above, we employed a mixed-methods approach to assess the system, including four self-reporting scales and open-ended questions. Given the compact workshop duration, we prioritized validated short-form scales to minimize participant fatigue.

\textbf{Momentary Hope Levels} was measured at both pre- and post-test intervals using the \textit{Children’s Hope Scale} \cite{snyder1997development}. This scale was selected to specifically evaluate changes in \textit{Agency} and \textit{Pathways}, the two core cognitive components underlying our theoretical framework. Post-intervention, we assessed \textbf{Narrative Immersion} using the \textit{Transportation Scale-Short Form (TS-SF)} \cite{appel2015transportation} to determine if the AI-scaffolded co-creation fostered deep cognitive absorption. We also measured \textbf{System Usability} via the \textit{UMUX-Lite} \cite{lewis2013umux} to verify that the prototype interface was accessible and appropriate for the target adolescent demographic. Finally, open-ended questions in the post-test solicited \textbf{Qualitative Feedback} regarding participants' subjective impressions of the collaborative storytelling activity.

\section{Preliminary Findings}

\begin{figure*}[htbp]
    \centering
    \begin{minipage}[b]{0.49\linewidth}
        \centering
        \includegraphics[width=\linewidth]{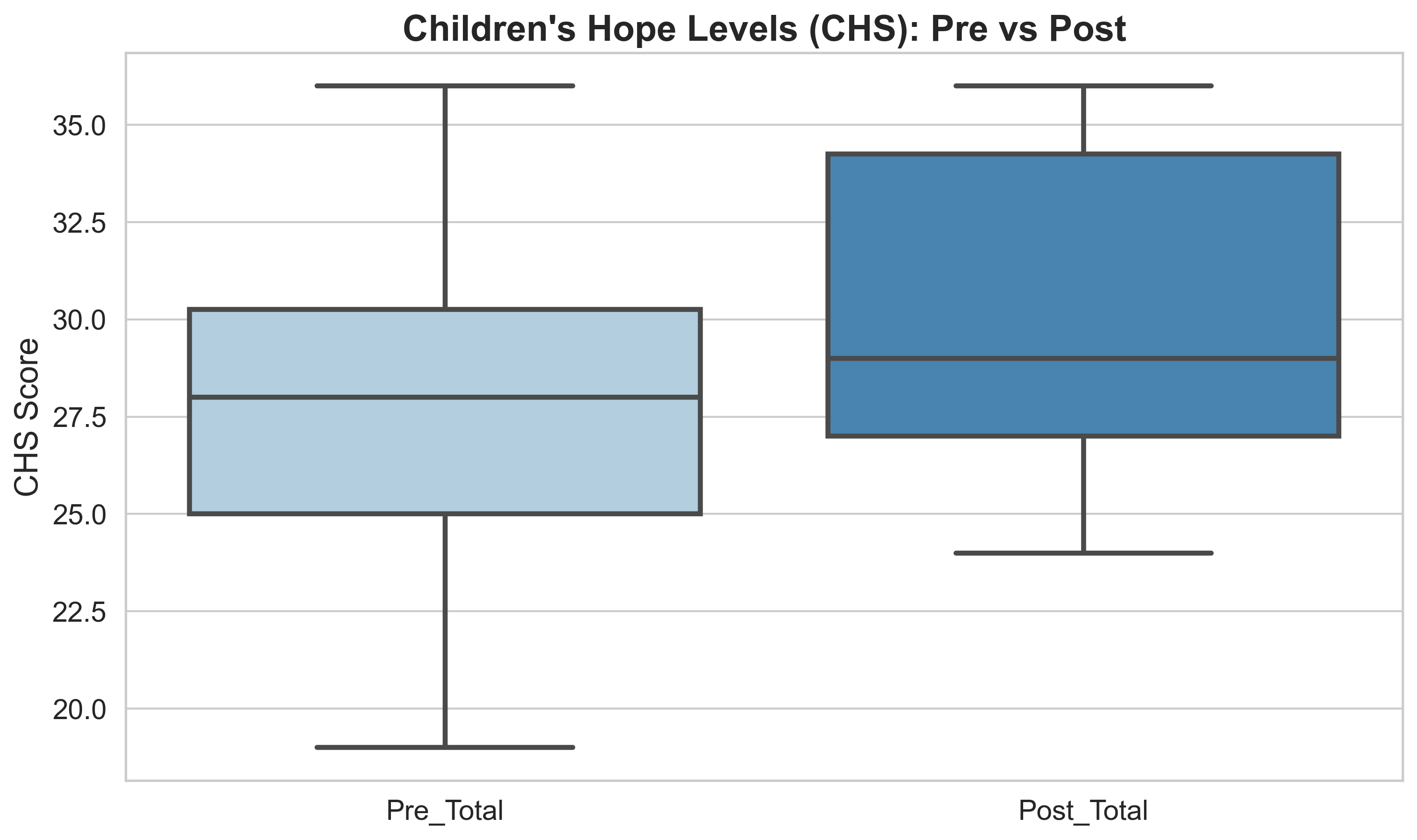}
        \centerline{\small (a) Pre- vs. Post-test CHS scores} 
    \end{minipage}
    \hfill 
    \begin{minipage}[b]{0.49\linewidth}
        \centering
        \includegraphics[width=\linewidth]{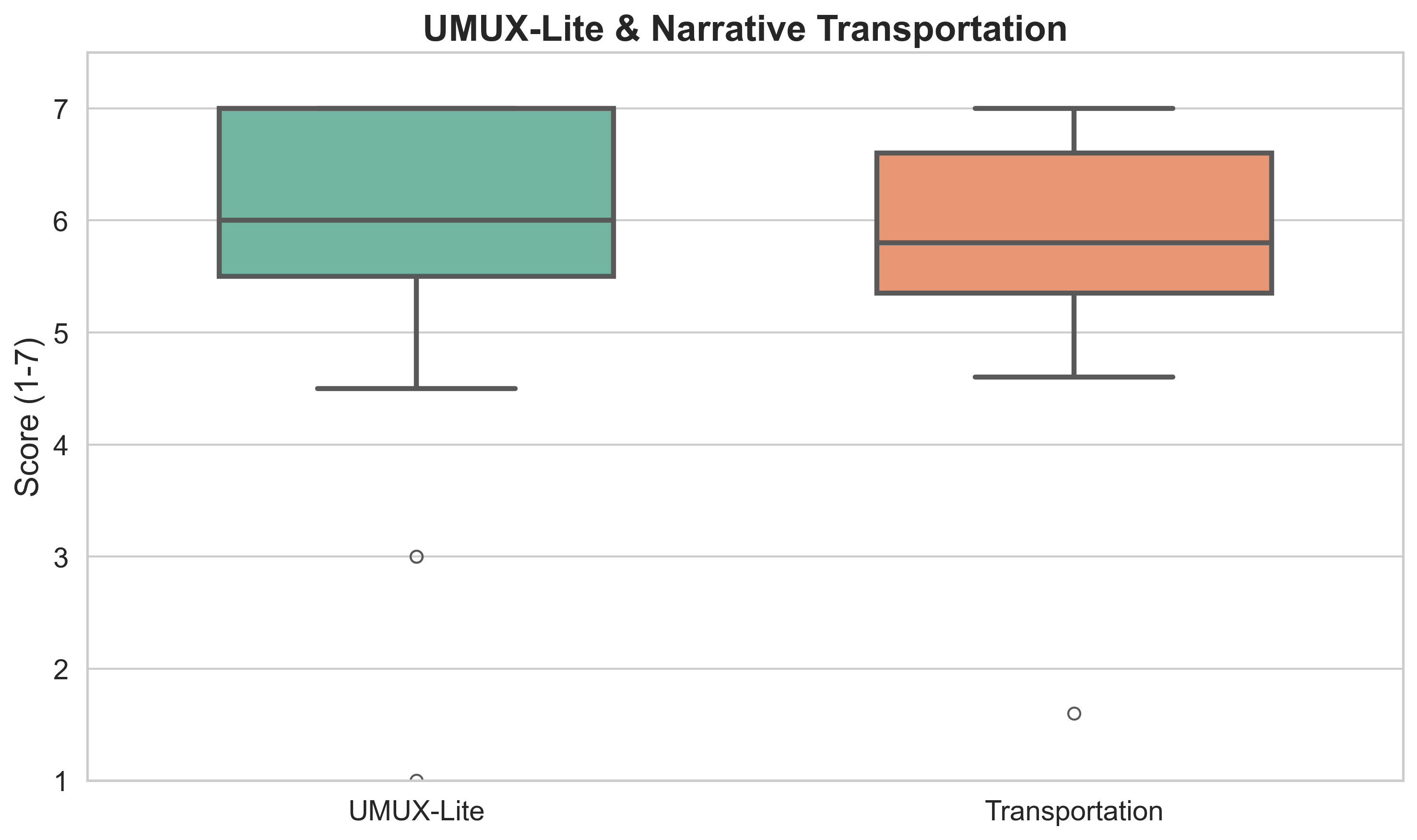}
        \centerline{\small (b) Usability and Immersion}
    \end{minipage}
    
    \caption{Quantitative evaluation results. (a) Boxplots showing a significant increase in hope levels scores ($p<0.05$). (b) Distribution of System Usability and Narrative Transportation scores, indicating positive user reception.}
    \label{fig:evaluation_results}
\end{figure*}


\subsection{Quantitative Results}
As illustrated in Figure \ref{fig:evaluation_results}(a), a paired sample t-test revealed a significant increase in total Hope scores from pre-test ($M=28.00$) to post-test ($M=30.00$), $t(19)=2.74, p=0.013, d=0.48$. 
Specifically, a breakdown of sub-scales indicated a significant improvement in \textit{Agency} ($M_{pre}=14.00, M_{post}=15.00, t(19)=2.15, p=0.044$), suggesting the workshop effectively boosted participants' goal-directed determination. 
However, the increase in \textit{Pathways} scores was marginally significant ($t(19)=1.95, p=0.066$). This implies that while students felt more empowered (Agency), their perceived capability to generate specific strategic routes may require longer-term scaffolding beyond a single session.


The Narrative Transportation Scale demonstrated excellent internal consistency (Cronbach's $\alpha = .90$). As shown in Figure \ref{fig:evaluation_results}(b), participants reported high levels of narrative immersion ($M=5.74/7, SD=1.25$), confirming effective engagement with the AI-generated scenarios. 
Regarding system usability, results from the UMUX-Lite were similarly positive ($M=5.72/7, SD=1.67$), with the distribution clustering at the upper end of the scale, validating the system's functionality and ease of use.


\subsection{Qualitative Reflections} 
Thematic analysis of the open-ended responses strongly corroborated the quantitative improvements in hope. Participants articulated a profound shift in their cognitive agency and resilience. For instance, P7-2 explicitly noted gaining \textit{higher expectations for the future''} and finding \textit{the power to strive,''} while P1-2 reflected that \textit{no dilemma is truly difficult''}—a sentiment indicating a strengthened belief in their capacity to navigate obstacles. This sense of empowerment was further echoed by P3-1, who reported enjoying \textit{the feeling of controlling the whole situation''} during the narrative construction. Beyond these primary agency-related outcomes, participants also acknowledged the system's role as a scaffold; feedback described the AI not only as an efficient tool for enriching narrative details (P3-3) but as a \textit{``human-like''} partner (P1-2) that fostered a creative and engaging team dynamic.

\section{Conclusion and Future Work}

This study presents a preliminary evaluation of \textit{FuturePrism}, demonstrating its potential to enhance adolescents' hope through AI-scaffolded collaborative storytelling. Quantitative results indicate a significant boost in \textit{Agency} alongside high immersion and usability, while qualitative feedback highlighted the empowering nature of the triadic AI partnership. Significantly, this work expands the scope of GenAI in education by moving beyond instrumental outcomes (e.g., writing proficiency) to pivot toward psychological well-being, representing one of the first explorations of AI-co-created narratives as an intervention for future thinking. While the marginal improvement in \textit{Pathways} suggests that developing strategic capabilities may require sustained engagement, these findings provide a promising proof-of-concept. Acknowledging limitations regarding sample size ($N=20$) and cultural context, future work will focus on rigorous longitudinal deployments to validate the sustainability of these psychological gains and their translation into behavioral changes across diverse settings.


\bibliographystyle{ACM-Reference-Format}
\bibliography{reference}

\end{document}